\documentclass[amsmath,amssymb,prd]{revtex4}
\usepackage{graphicx}% Include figure files
\usepackage{dcolumn}% Align table columns on decimal point
\usepackage{bm}% bold math

\begin{document}

%\preprint{APS/123-QED}

\title{Efficient photon sorter in a high-dimensional Hilbert space}

\author{Warner A. Miller}
\email{wam@fau.edu}
%\homepage{http://physics.fau.edu/~wam}
 \affiliation{Department of Physics, Florida Atlantic University,  Boca Raton, Florida 33431}

%\date{December 6 ,2010}

\begin{abstract}
An increase in the dimension of state space for quantum key distribution (QKD) can decrease its fidelity requirements while also increasing its bandwidth. A significant obstacle for QKD with qu$d$its  ($d\geq 3$) has been an efficient and practical quantum state sorter for photons whose complex fields are modulated in both amplitude and phase.  We propose such a sorter based on a multiplexed thick hologram, constructed e.g.  from photo-thermal refractive (PTR) glass. We validate this approach using coupled-mode theory with parameters consistent with PTR glass to simulate a holographic sorter.  The model assumes a three-dimensional state space spanned by three tilted planewaves.  The utility of such a sorter for broader quantum information processing applications can be substantial.
\end{abstract}

\maketitle

\section{QKD with  qu$d$its}        
We concern ourselves here with the secure distribution of a one time
only key from a sender (Alice) to a receiver (Bob). The three elements
of any quantum key distribution (QKD) system are: (1) Alice must be
able to prepare at will a single photon state chosen from a set of
mutually-unbiased bases (MUB) \cite{Schwinger:60,Wootters:89}, (2)
each of these quantum amplitudes must be propagated from Alice to Bob
with reasonable fidelity, and finally (3) Bob must have the ability to
choose between one or another of the MUBs and, if he chooses the
correct basis, be able to efficiently sort and detect each of these
photon states.  This QKD scenario has been exhaustively studied in the
literature and is replete with security proofs for numerous protocols,
e.g.  \cite{Bennett:84,Ekert:91,Ekert:00}. These security proofs have
been extended in many cases to higher-dimensional state spaces
\cite{Cerf:02}, and all of the protocols have been or are currently
being demonstrated successfully \cite{Groblacher:06, Howell:06}.

One of the conventional realizations of QKD today involve transmitting
heavily-attenuated laser pulses from Alice to Bob and encoding qubit
information in each packet by utilizing the spin of the photon. This
allows Alice and Bob, who are suitably authenticated, the possibility
to establish and share an arbitrarily-secure one-time only key between
them. Here they have access to a two-dimensional state space and can
therefore form three MUBs each with two orthogonal polarization
states. Such a six-state QKD scheme \cite{Bruss:98,Enzer:02} has limited
bandwidth and optical fidelity constraints. These constraints can be
ameliorated by extending the QKD to higher-dimensional state space
\cite{Cerf:02}.

The potential of extending photon-based QKD to higher dimensions was
made possible in 1983 when Miller and Wheeler \cite{Miller:84} described a fundamental 
quantum experiment utilizing a photon's orbital angular momentum (OAM).    In 1992 Allen et~al.\cite{Allen:92}  described Laguerre-Gaussian light beams that possessed a quantized orbital angular momentum (OAM) of $l\hbar$ per photon.   This opened up an arbitrarily high
dimensional quantum space to a single photon \cite{Miller:84,Allen:03}.
Following these discoveries Mair et~al. \cite{Mair:01,Oemrawsingh:04}
unequivocally demonstrated the quantum nature of photon OAM by showing
that pairs of OAM photons can be entangled using parametric down
conversion. Shortly thereafter, Molina-Terriza
et~al. \cite{Molina-Terriza:02} introduced a scheme to prepare photons
in multidimensional vector states of OAM commencing OAM QKD.

While photons with specific values of OAM have been emphasized in the
literature we can equally well utilize any other set of orthogonal
basis functions for higher-dimensional QKD.  While OAM states respect
azimuthal symmetry, linear momentum (LM) states respect rectilinear
symmetry.  Independent of the representation we use, the MUB states
will ordinarily be modulated in both amplitude and phase.  Recently a
practical method has been demonstrated to produce such MUB states
using computer-generated holography with a single spatial light
modulator (SLM) \cite{Gruneisen:08}.

While the advantage of higher-dimensional QKD lies in its ability to
increase bandwidth while simultaneously tolerating a higher bit error
rate (BER) \cite{Cerf:02}, Two potential problems confront this
approach. First, such transverse photon wave functions are more
fragile in propagation than the photon's spin
\cite{Paterson:05,Aksenov:08}, and the divergence of the states
($\propto\! \sqrt{l}$) may require larger apertures. Despite this,
multi-conjugate adaptive optical communication channels may be able to
ameliorate these problems \cite{Paterson:05b,Tyler:08}.  A second
obstacle involves the efficient sorting of qu$d$it MUB-state photons
with small Fock-state quantum numbers, and the paper will address this
particular problem.  Currently, the only solution to this problem for
the case of OAM photons is the use if a cascaded Mach-Zehnder
interferometric system \cite{Xue:01,Leach:04,Zou:05}. Proposed systems
of this kind have been demonstrated only for 4-dimensions and can be exceedingly difficult to stabilize.  
Other approaches that use crossed thin
diffraction gratings are not efficient enough to establish a secure
key.  Furthermore, the design of a practical and efficient sorter for photons 
with OAM remains elusive; however, we argue here that such sorters 
can be constructed for photons with LM within a single optical element.  It
is for these reasons that we concentrate on LM in this manuscript. 

We focus here on the efficient sorting of single photons with
arbitrary complex wavefronts. Ideally, what is needed is an qu$d$it
version of a polarization sorter, i.e. a single optical element, one per MUB
basis, that can efficiently sort each of the qu$d$it states in that
basis while equally distributing every other qu$d$it state. Thick
holographic gratings fortunately produce high diffraction efficiency
in the first order \cite{Goodman:05,Kogelnik:69,Case:75}. If several
predominant diffracted orders are required, as is the case for
sorting, several independent fringe structures can exist in the
emulsion.  Such multiplexed holograms have been used for multiple-beam
splitters and recombiners \cite{Case:75} and more recently for
wide-angle beam steering\cite{Ciapurin:06}.  In this paper we propose
such a MUB-state sorter based on a multiplexed thick holographic
element constructed from commercially available photo thermal
refractive (PTR) glass \cite{Efimov:99}. Due to the unique properties
of PTR glass the grating's thickness can approach several {\it mm} and
be highly Bragg selective. There is evidence that such sorters can be
highly efficient, $> 95\%$ \cite{Ciapurin:06}. Our simulations
presented here and empirical data on thick Bragg gratings indicate
that they may provide an adequate solution to this critical problem.

Before we describe our proposed thick holographic MUB sorter we will
briefly focus our attention on ``twelve-state QKD'' and work in a
$3$-dimensional state space. This is one more dimension than that
available to photon polarization states and serves to illustrate our
approach. Nevertheless, our work is equally applicable to higher
dimensional state spaces.  Its limitations will require further
investigation.

\section{Linear momentum QKD in a 3-dimensional state space}
\noindent

In 3-dimensions there are a maximum of four MUBs which we refer
to here as $MUB_1$, $MUB_2$, $MUB_3$ and $MUB_4$.  Each of these
orthonormal bases contain three state vectors.  Therefore there are twelve quantum
states at our disposal for QKD.  If we identify
$|a\rangle$, $|b\rangle$ and $|c\rangle$ as the orthonormal ket
vectors of $MUB_1$, then the other nine qu$d$it states from the other
three MUB bases are specific linear combinations of these
(Table~1).
%\vspace*{4pt}   %only when needed
\begin{table}[hb]
\label{table:1}
\caption{The four MUBs in our 3-dimensional state space. Note that
  phase, $\xi=\exp(i2\pi/3)$,  is a cube root of unity, and we have suppressed the
  normalizing factor of $1/\sqrt{3}$ in each of the nine MUB states in
  the last three columns.}
%\centerline{\footnotesize}
%\centerline{\footnotesize\smalllineskip}
\begin{center}
\begin{tabular} {| c |c | c | } \hline
$MUB_1$ & $MUB_2$  \\ \hline
\( \begin{array}{ll}
|a \rangle \\
|b \rangle \\ 
|c \rangle
\end{array} \) &
\( \begin{array}{lll}
|a_2 \rangle \propto  | a \rangle + &\!\!\! | b \rangle + &\!\!\! | c \rangle \\
|b_2 \rangle \propto  | a \rangle + \xi &\!\!\! | b \rangle + \xi^2 &\!\!\! | c \rangle \\
|c_2 \rangle \propto  | a \rangle + \xi^2 &\!\!\! | b \rangle +  \xi &\!\!\! | c \rangle 
\end{array} \) \\ \hline
\end{tabular}

\begin{tabular} {| c | c | } \hline
$MUB_3$ & $MUB_4$\\ \hline
\( \begin{array}{lll}
|a_3 \rangle \propto  | a \rangle + &\!\!\! | b \rangle + \xi &\!\!\! | c \rangle \\
|b_3 \rangle \propto  | a \rangle + \xi &\!\!\! | b \rangle +  &\!\!\! | c \rangle \\
|c_3 \rangle \propto  | a \rangle + \xi^2 &\!\!\! | b \rangle +  \xi^2 &\!\!\! | c \rangle 
\end{array} \)  &
\( \begin{array}{lll}
|a_4 \rangle \propto   | a \rangle + &\!\!\! | b \rangle + \xi^2 &\!\!\! | c \rangle \\
|b_4 \rangle \propto   | a \rangle + \xi &\!\!\! | b \rangle + \xi &\!\!\! | c \rangle \\
|c_4 \rangle \propto  | a \rangle + \xi^2 &\!\!\! | b \rangle + &\!\!\! | c \rangle \\ 
\end{array} \) \\ \hline
\end{tabular}
\end{center}
\end{table}

For the purpose of this paper we can simplify our analysis and retain
the physical content by quantizing in the space of linear momentum
(LM-QKD) rather than in angular momentum (OAM-QKD). Furthermore, without loss of generality
we can concentrate in this section on the physics of LM in only one transverse dimension.  As a result we can
freely choose as $MUB_1$ any three non co-linear planewaves. In this
case, our three integer quantum numbers will be the number of waves of
tilt of these planewaves with respect to the normal of the holographic
emulsion across an aperture of breadth $D$. Here we assume the hologram surface is in the $x$-$y$ plane.  These waves correspond to a transverse linear
momentum $p^x_a = a \hbar k_\perp$, $p^x_b = b \hbar k_\perp$ and $p^x_c = c
\hbar k_\perp$; respectively.  Here,  $k_\perp = k \lambda/D$ is the magnitude of the 
x-component of the wave vector of a plane wave with one wave of tilt,
$\tau\! \sim \lambda/D$, across the aperture.

In the frame of the hologram and in units where the speed of light is
unity, the components of the 4-momentum, $p = \{ p^t,p^x, p^y, p^z
\}$, of each of our three photons can be expressed in terms of their
transverse momentum, $k_\perp$ and wavenumber, ($k$), i.e.
\begin{eqnarray*}
p_{a} &=  \hbar k_a &= a \hbar k_\perp \ \left\{(k/ak_\perp)^2,1,0,\sqrt{(k/a k_\perp)^2-1}\right\},\\ 
p_{b} & = \hbar k_b &= b \hbar k_\perp \ \left\{(k/bk_\perp)^2,1,0,\sqrt{(k/b k_\perp)^2-1}\right\},\   \hbox{and}\\
p_{c} &=  \hbar k_c &= c \hbar k_\perp \ \left\{(k/ck_\perp)^2,1,0,\sqrt{(k/c k_\perp)^2-1}\right\}. 
\end{eqnarray*}
In the remainder of this paper the transverse linear
momentum wavenumbers represent our three quantum numbers for
k-QKD. These three planewaves define our first MUB and is given by,
\[
MUB_1 = \left\{ | a \rangle, | b \rangle, | c \rangle \right\}.
\]
Each of these states represents a transverse Fourier mode of a photon;
they are orthogonal ($\langle i | j \rangle = \delta_{i,j}$) and
define our 3-dimensional state space.  The other nine MUB states
(Table~1) can be obtained from these by linear
superposition and will represent wavefronts with both amplitude and
phase variations.

Just as the wavefront of an optical vortex ($\propto \exp{i l \phi}\exp{i k z}$) has a beam divergence in $z$ proportional to $l$,  the LM  states also have a beam divergence proportional to the number of waves of tilt.  These divergences  can be accommodated in free-space communication links with appropriate telescope apertures.  In this sense the LM states share this limitation with photon OAM states.  However, one can quickly determine that communication links between satellite and ground for near-earth orbits can be achieved with reasonably-sized telescopes.  Therefore, the utility of  LM states  for quantum communication and quantum information processing we expect to be  similar to that of OAM states. Given their linearity and simplicity,  LM states may have advantage in some situations. 

\begin{figure} [htbp]
%\vspace*{13pt}
%\centerline{\epsfig{file=figure1.eps, width=8.2cm}} %100 percent
\begin{center}
\includegraphics[width=0.5\linewidth]{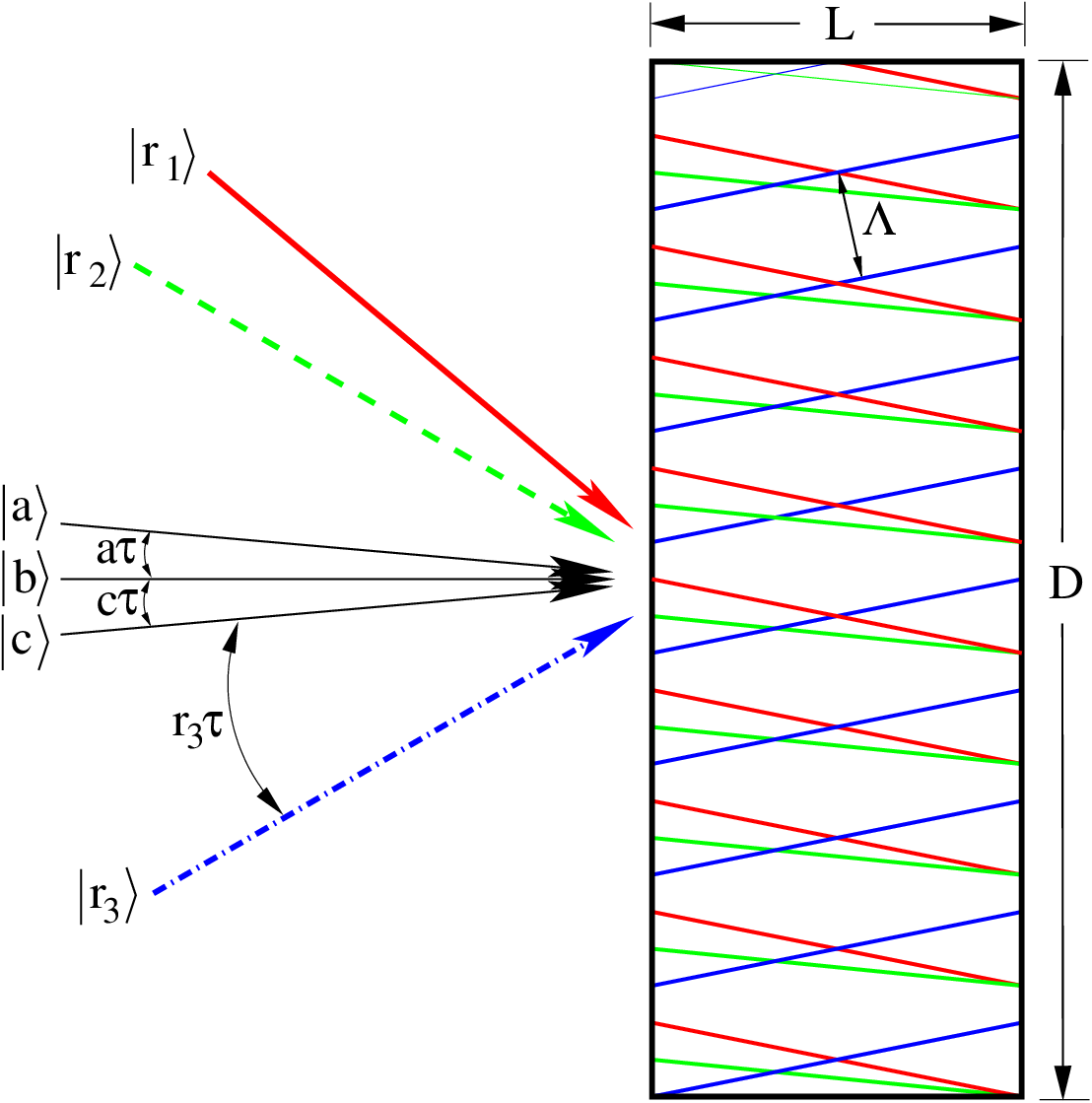}
\end{center}
\vspace*{13pt}
\caption{\label{fig:vh} We provide here an illustration of our proposed thick holographic $MUB_4$
  sorter (the other three sorters will be of similar design). The three signal waves are the appropriate linear
  superpositions of the three qu$d$it states of $MUB_4$ shown in
  Table~1.}
\end{figure}

For each MUB we consider a multiplexed thick holographic sorter,
i.~e. a triple-exposed grating structure formed by the incoherent
superposition of three gratings within a single emulsion. Here, each
grating is formed by the superposition of the respective MUB state and
its own unique plane reference wave.  In this paper we concentrate on
the construction of the $MUB_4$ sorter (Fig.~\ref{fig:vh}), as the
other three MUB sorters will be of similar design.

We construct the $MUB_4$ sorter in three steps. First, we record the
interference pattern of our first signal wave $| a_4 \rangle$ with a
corresponding reference planewave, $| r_1\rangle$ having $r_1\! \gg\! 1$
waves of tilt. After this initial recording is complete, we then
record the interference pattern of our second signal state from
$MUB_4$, namely $| b_4 \rangle$ with a second reference planewave
$ | r_2\rangle$. Finally, we record a third independent set of
fringe patterns by interfering the qu$d$it signal state $| c_4
\rangle$ with a third reference planewave $| r_3 \rangle$.  This
produces a triple-multiplexed hologram.

We show that the hologram described above faithfully represents a quantum
projection (``redirection'') operator for $MUB_4$, where
\begin{equation}
\label{po}
{\cal P}_4 = | r_1 \rangle\langle a_4 | + | r_2 \rangle\langle b_4 | + 
             | r_3 \rangle\langle c_4 |.
\end{equation}
Its operation on any one of the 12 MUB states should produce the
desired result.  In other words, if the hologram is illuminated by MUB
state $| a_4 \rangle$, $| b_4 \rangle$ or $| c_4 \rangle$ it should
produce a planewave in state  $| r_1 \rangle$, $| r_2 \rangle$
or $| r_3 \rangle$, respectively. If it is illuminated by any of the
other nine MUB states it should then produce an equally weighted response
into all three reference states, e.g. 
\[
\langle r_i | {\cal P}_4 | b_3 \rangle =  \frac{1}{\sqrt{3}},\ \forall i\in\{1,2,3\}.
\]

The unique property of PTR glass with its bulk index of refraction,
$n_0=1.4865$, and depth of modulation, $\Delta n/n_0= 336\, ppm$,
place it squarely in the realms of scalar diffraction theory and
coupled-mode (CM) theory. Furthermore, a PTR hologram can be thick,
$L\! \sim\!  D\sim\! 1\, cm$, with Bragg-plane periods, $\Lambda \!
\sim\!  \lambda$. Consequently, their Bragg selectivity, $\sim \!
\Lambda/L$, can approach the diffraction limit of one wave of tilt
across its aperture \cite{Ciapurin:06}.  A wavelength,  $\lambda\!
=\! 1085\, nm$,  and an aperature,  $D\! \sim\!  L\sim 1\, cm$,  yield a minimal divergence of
our three signal waves of a few $arcsec$.

\section{Coupled mode analysis of the LM holographic sorter}
\label{sec:CM}
\noindent
To examine the $MUB_4$ sorter we will follow closely the CM approach of
Kogelnik \cite{Kogelnik:69} and the notation used by Case \cite{Case:75}
to solve the scalar wave equation, 
\begin{equation}
\label{swe}
\nabla^2 E_y + k^2 E_y = 0,
\end{equation}
 for polarization perpendicular to the plane of incidence.
Here, the linearly-polarized electric field, $E_y(x,z)$,  of frequency,
$\nu$, is assumed to be independent of $y$. Following CM
theory we keep only primary modes for the electric field. These are
the transverse harmonic modes given by the k-vectors $\vec k_1$, $\vec
k_2$, $\vec k_3$, $\vec k_a$, $\vec k_b$ and $\vec k_c$ associated to
planewave reference states $|r_1\rangle$, $|r_2\rangle$, $|r_3\rangle$
and $MUB_1$ signal states $|a\rangle$, $|b\rangle$, $|c\rangle$,
respectively. Hence, 
\begin{eqnarray*}
E_y (x,z) & = R_1(z) \exp^{\vec k_1 \cdot \vec r} + 
            R_2(z) \exp^{\vec k_2 \cdot \vec r} + 
            R_3(z) \exp^{\vec k_3 \cdot \vec r}  \\ 
          & +   S_a(z) \exp^{\vec k_a \cdot \vec r} + 
            S_b(z) \exp^{\vec k_b \cdot \vec r} + 
            S_c(z) \exp^{\vec k_c \cdot \vec r}.  
\end{eqnarray*}
Here, the six mode amplitudes, $\{R_i\}$ and $\{S_i\}$ are only functions of $z$.  They are set initially to $\{R_i(z=0)\} =\{1,1,1\}$ and to the corresponding amplitude and phase factors of one of the twelve corresponding signal states shown in Table~1. For example, given the signal state $|c_4>$,  one would set $\{S_i\} = 1/\sqrt{3} \{ 1,\xi^2,1\}$. The wavenumber $k(x,z)$ of Eq.~\ref{swe} represents the three incoherently recorded gratings mentioned above,
\[
k = n(x,z) k_0  = \underbrace{n_0 k_0}_{\beta} \left( 1 + \frac{\Delta n}{n_0}
\frac{\left(I_{R1} + I_{R2} + I_{R3}\right)}{6(1+\sqrt{3})} \right),
\]
where $I_{Ri}$ is the intensity modulation of the $i^{th}$ grating,  e.g.,
\[
I_{R3} = 2 - |e^{ik_3\cdot r} + \frac{1}{\sqrt{3}}\left( e^{ik_a\cdot r} + \xi\, e^{ik_b\cdot r}+ e^{ik_c\cdot r}
\right) |^2.
\]
Following the usual approximations of CM theory \cite{Kogelnik:69, Goodman:05},  we assume  that the amplitude functions,  $\{R_i\}$ and $\{S_i\}$ are slowly varying  functions of $z$ and we can neglect the second derivative terms, yielding the six equations for the mode amplitudes,
%\begin{widetext}
\begin{equation}\label{CM}
\left( \begin{array}{c} 
                        R'_1 \\
                        R'_2 \\
                        R'_3 \\
                        S'_a \\
                        S'_b \\
                        S'_c
       \end{array} \right) = i \kappa^2
\left( \begin{array}{c c c c c c } 
                       0 & 0 & 0 & \frac{1}{\rho_1} & \frac{1}{\rho_1}    & \frac{z}{\rho_1}   \\
                       0 & 0 & 0 & \frac{1}{\rho_2} & \frac{z^*}{\rho_2}  & \frac{z^*}{\rho_2} \\
                       0 & 0 & 0 & \frac{1}{\rho_3} & \frac{z}{\rho_3}    & \frac{1}{\rho_3}   \\
                       \frac{1}{\sigma_a} & \frac{1}{\sigma_a}    & \frac{1}{\sigma_a} & 0 & 0 & 0   \\
                       \frac{1}{\sigma_b} & \frac{z}{\sigma_b}    & \frac{z^*}{\sigma_b} & 0 & 0 & 0   \\
                       \frac{z^*}{\sigma_c} & \frac{z}{\sigma_c}    & \frac{1}{\sigma_c} & 0 & 0 & 0 
        \end{array} \right)
\left( \begin{array}{c} 
                        R_1 \\
                        R_2 \\
                        R_3 \\
                        S_a \\
                        S_b \\
                        S_c
       \end{array} \right),
\end{equation}
%\end{widetext}
where $\kappa^2 \equiv \left(\frac{\beta^2}{6(3+\sqrt{3})}\right)
\left( \frac{\Delta n}{n}\right)$, and $\rho_i = k^z_i$ and
$\sigma_j=k^z_j$ are the $z$-components of the wave vectors for the
reference and signal states, respectively.  

\section{A recording geometry yielding perfect efficiency}
\begin{figure} [htbp]
%\vspace*{13pt}
%\centerline{\epsfig{file=figure2.eps, width=8.2cm}} %100 percent
\begin{center}
\includegraphics[width=0.6\linewidth]{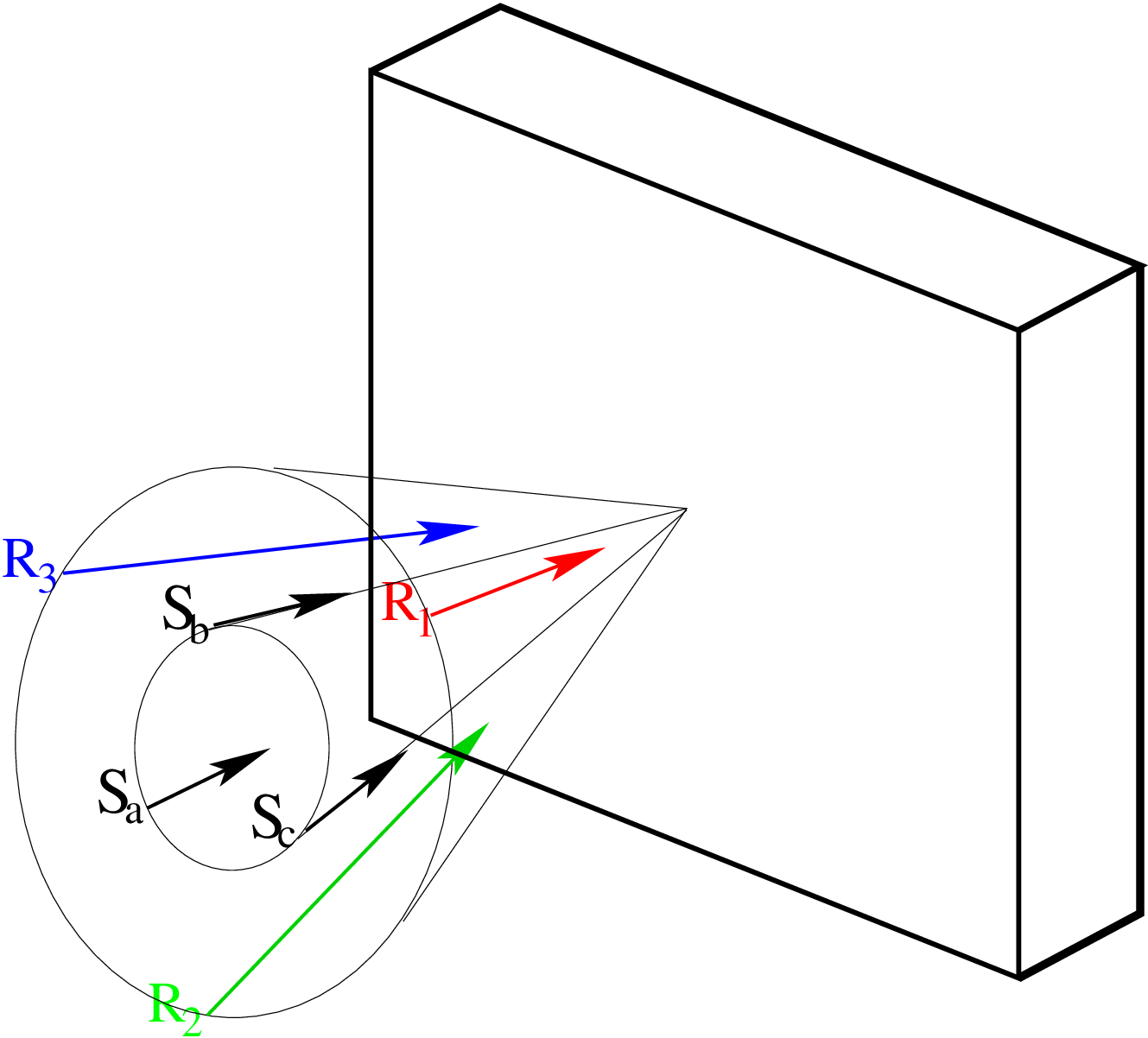}
\end{center}
\vspace*{13pt}
\caption{\label{fig:cyl}We illustrate the recording geometry of a proposed thick holographic $MUB_4$ sorter (the other three sorters will be similarly recorded) that can achieve perfect efficiency within the assumptions of CM theory. The three signal waves are in the appropriate linear superpositions of the three qu$d$it states of $MUB_4$ as shown in Table~1.}
\end{figure}

\noindent
It is possible within the assumptions of CM theory to record a multiplexed hologram for which the solutions of Eq.~\ref{CM} yield an perfectly efficient sorter.  Ordinarily the efficiency of a single volume Bragg grating is an oscillatory function of depth ($z$) into the hologram. Its period depends on the grating strength. In our notation the efficiency given by, 
\begin{equation}
\left( \begin{array}{c} 
\hbox{\em Diffraction Efficiency of a}\\ 
\hbox{\em Volume Bragg Grating}
\end{array}\right)  = sin{\left( \frac{\kappa^2}{\sqrt{\rho\sigma}}\ z\right)},
\end{equation}  
achieves its first maxima at, 
\begin{equation}
L= z_{max} =  \frac{1}{2} \left(\frac{n}{\Delta n}\right) \lambda \frac{\sqrt{\rho\sigma}}{n k}.
\end{equation}
For optimal efficiency the Bragg grating must be tuned to this depth.  However, this depends on the square root of the products of the $z$-components of both the signal ($\sigma$)  and reference ($\rho$) waves. Therefore, for a given multiplexed MUB hologram (e.g. $MUB_4$),  we would need to assure that the maxima for each of the three recordings were equal to each other.  Fortunately one class of configurations satisfies this condition.  In particular if we 
demand (1) that all  signal $k$-vectors are equal,  $\sigma_a=\sigma_b=\sigma_c=\sigma$,  and   (2) that all reference k-vectors are the same,  $\rho_1=\rho_2=\rho_3=\rho$, then we can solve Eq.~3 analytically (Fig.~\ref{fig:cyl}).   In other words, each of the three signal  $k$-vectors lie on the surface of a cone centered about the normal to the hologram's surface  and subtended by an angle $\theta_s= a \lambda/D$ for some positive quantum number  $a $. Similarly, each of the the three reference $k$-vectors lie on a second concentric cone subtended by angle $\theta_r$, where $\rho=k \cos \theta_r$. 

\begin{figure} [htbp]
\label{fig:MUB4}
\begin{center}
\includegraphics[width=0.65\linewidth]{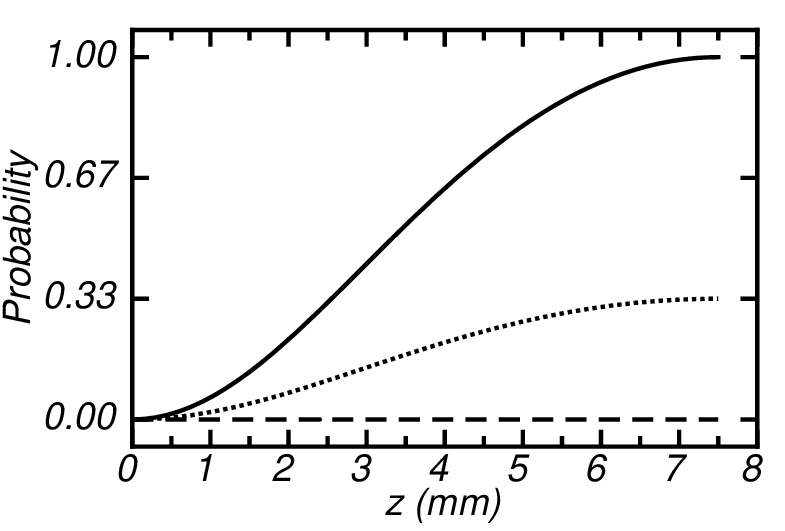}
\end{center}
\vspace*{13pt}
\caption{Here we show the predicted CM performance for the multiplexed  $MUB_4$ sorter.   This single optical element is a triple multiplexed volume hologram constructed using PTR glass ($n_0=1.4865$, $\Delta n=0.0005$ at $\lambda = 1085\, nm$).   The three reference signals are recorded on a cone (Fig.~2) with reference angle, $\theta_r=\pi/4$,  and the three signal waves on a cone with angle, $\theta_s=\pi/90$.  The small ratio of depth of modulation to bulk index ($\Delta n/n_0 \sim 336 ppm$) places it squarely in the linear regime of CM theory.   The graph shows the  probability that the signal photon will be observed to be diffracted into one or another of the three original reference directions ($\langle \vec r | r_1 \rangle=R_1 e^{i k_1 \cdot r}$,
  $\langle \vec r | r_2 \rangle = R_2 e^{i k_2 \cdot r}$ and  $\langle
  \vec r | r_3 \rangle = R_3 e^{i k_3 \cdot r}$) as a function of the
  depth ($z$) of the emulsion. The curves have been terminated at
  their maximum efficiency depth ($z_{max}\approx 7.5\,mm$).      If we send one of the 12 $MUB$ states (Table~1) into this  $MUB_4$ sorter we can expect one of three characteristic diffraction patters (solid, dotted and dashed lines).  First, for any photon prepared in any one of the 9 qudit  states from $MUB_1$, $MUB_2$ or $MUB_3$  there is equal probability that the photon will be diffracted into the three directions  ($r_1$, $r_2$ and $r_3$) thereby yielding no information as to identity of the quantum state (dotted curve). However, if the signal is one of the three  $MUB_4$ states it will be diffracted with certainty into its corresponding reference wave direction (solid curve) and not at all into the other two channels (dashed curve).  This yields complete information as to the identity of the quantum state.}
\end{figure}

In solving Eq.~\ref{CM} for $MUB_4$ with signal state $|a_4\rangle$,  we find the probability for the photon to be diffracted into the correct detector sensitive to $|r_4\rangle$  to be a cyclic function  in $z$,
\begin{equation}
Probability =  \left( 
\frac{ \sigma }{ \sigma + \rho \cot^2 
\left(  
\frac{\sqrt{3}\, \kappa^2}{\sqrt{\rho\,\sigma}} z 
\right)} 
\right).
\end{equation}
The first maximum corresponds to perfect efficiency ($Probability = 1$)  and occurs at,
\begin{equation}\label{zmax}
L=z_{max} = \frac{\sqrt{3} \pi\, \sqrt{\rho \sigma}}{6\, \kappa^2}.
\end{equation}
Furthermore, the solution to Eq.~\ref{CM} for this recording geometry
and for each of the twelve initial signal MUB  states
are shown in Fig.~\ref{fig:MUB4}.   This geometry faithfully reproduces the desired
projection operator, Eq.~\ref{po}.   As a further verification of the assumptions inherent in the CM analysis, we independently examined the
far-field pattern for such gratings using a finite difference
time domain solution of Maxwell's equations and observed that the CW
assumptions are valid.  Only the primary modes were dominant, and our results are consistent with scalar diffraction theory so that  there were no relevant polarization changes observed in the field.

\section{Discussion on the limitations of the grating selectivity  for OAM and LM photons}
\noindent
A necessary condition for our gratings to be able to efficiently sort photons with OAM is that the angular (``Bragg'')  selectivity of the volume hologram should be at least sensitive enough to differentiate such wave front tilts.  This  places constrains on the design  of our grating given the value of OAM and the aperture of the beam.  The most stringent condition ($\Delta \phi_1 \le \Delta \phi_l$, $\forall l\ge1$) occurs when there is the smallest wavefront tilt across the aperture. In the case of an OAM photon, the  $l=1$ state is the state with the smallest pitched helical wavefront.  Here we would require that,
\begin{equation}
\label{condition}
\Delta \theta_{Bragg} \le \Delta \phi_1.
\end{equation}

We can estimate this wavefront tilt for three relevant cases, (1) for LM, (2) for an optical vortex, and (3) for a Laguarre-Gaussian (LG) photon. First, the wavefront tilt  for pure-state LM photons with wave vector $k$  is constant across the aperture of breadth $D$ and is given by the quantum number $b$,
\begin{equation}
\label{lmt}
\Delta \phi_{LM} \sim \frac{2 \pi b}{k D} .
\end{equation}
Secondly, for an optical vortex the tilt of the wavefront decreases inversely with radius.  The tilt  at the boundary of the aperture is,
\begin{equation}
\label{ovt}
\Delta \phi_{OV} \sim \frac{|l|}{k D}.
\end{equation}
We now estimate the tilt of a Laguerre-Gaussian beam of amplitude $A_l$ with angular momentum $l$ and  zero radial index,  
\begin{equation}
\label{LGE}
E_{l}(\rho,\phi,z) =  A_{m} \left( \frac{w_0}{w(z)}\right) \left( \frac{\rho}{w(z)} \right)^l e^{-\rho^2/w^2(z)} L^l_0\left( \frac{2 \rho^2}{w^2(z)} \right) 
e^{i(l+1)\gamma(z)-i k  \rho^2/2R(z)}e^{il\phi} e^{ik z}.  
\end{equation}
Here,  the beam of minimum waist,$w_0$,  has Gouy phase, $\gamma(z) = \arctan{\left(z/z_R\right)}$, a wavefront curvature, 
\begin{equation}
R(z) = z\left( 1+ \left(z_R/z\right)^2\right), 
\end{equation}
and a waist,
\begin{equation}
w(z) = w_0 \sqrt{1 + \left( \frac{z}{z_R}\right)^2}, 
\end{equation}
that is scaled to the Raleigh range, $z_R= \pi w_0^2/\lambda$.
 The tilt of the Poynting vector of this helical wavefront along the circumference of its maximum intensity ($r_{max}$) is a function of its waist, wave number and angular momentum,
\begin{equation}
\delta \phi_l= \frac{|l|}{2 k r_{max}}.
\end{equation}
This skewness of the wavefront has been observed recently using a Shack-Hartmann wavefront sensor \cite{L06}. 
For a LG photon (Eq.~\ref{LGE}) the location in radius  of the annulus of maximum intensity is, 
\begin{equation}
r_{max} = \sqrt{\frac{|l|}{2}}\, w(z). 
\end{equation}
Therefore, we require a grating with angular selectivity, 
\begin{equation}
\label{lgt}
\Delta \phi_{LG} \le  \sqrt{\frac{|l|}{2}}\, \frac{1}{k w(z)}.
\end{equation}
 
While we do not have a theory for the angular selectivity for our OAM or multiplexed LM gratings we {\em assume} here that the angular selectivity is similar to the corresponding ($l=0$) linear Bragg grating -- a selectivity that is analytically tractable. Perhaps this  a good motivation for the further development of a numerical wavefront propagator for volume holograms.  

For a linear Bragg grating the efficiency, $\eta$, as a function of small mismatch in the illumination away from the Bragg angle ($\theta' = \theta_B - \Delta \theta$) is easily calculated using CM theory \cite{Kogelnik:69,Ciapurin:06,Goodman:05},   
\begin{equation}
\eta = \frac{ \sin^2{\left( \Phi \sqrt{1+\chi^2/\Phi^2}\right)}}{1+\chi^2/\Phi^2}.
\end{equation}
Here $\Phi = \pi/2$ which means that the thickness, $L$,  of the emulsion is tuned for maximum efficiency, with 
$L/\cos{(\theta_B)} = \lambda/2\Delta n$, and $\theta_B$ is the incident angle of the reference beam, where $\cos{(\theta_B)} = (\vec k_r \cdot \hat z)/|k_r|$.  The variable,
\begin{equation}
 \chi= \frac{\xi\, L}{2 \cos{\theta_B}}, 
 \end{equation}
 is related to the angular detuning parameter, 
 \begin{equation}
 \label{dp}
 \xi = K \left( \Delta \theta \cos{(\theta_B - \psi)} \right), 
 \end{equation}
where $K= 2\pi/\Lambda$ is the wavenumber of the grating planes and $\psi$ is angle between the grating planes and the normal to the emulsion surface. 

Here we will define our angular selectivity ($\Delta \theta_{Bragg}$)  to be twice the the mismatch angle that gives half efficiency, $\eta = 1/2$, i.e. 
\begin{equation}
\Delta \theta_{Bragg} = 2 \Delta \theta_{1/2}.
\end{equation} 
First we solve,
\begin{equation}
\frac{1}{2} = \frac{ \sin^2{\left( \frac{\pi}{2} \sqrt{1+4 \chi_{1/2}^2/\pi^2}\right)}}{1+4 \chi_{1/2}^2/\pi^2},
\end{equation}
for $\chi_{1/2}$.  This occurs when 
\begin{equation}
\chi_{1/2} \approx 1.247  \approx \frac{5}{4},
\end{equation}
or when the detuning parameter, 
\begin{equation}
\xi_{1/2} \approx \frac{5 \cos{\theta_B}}{2 L},
\end{equation}
which when substituted into Eq.~\ref{dp} gives the angular sensitivity of our volume phase grating,
\begin{equation}
\Delta \theta_{Bragg} \approx  5 \left( \frac{ \cos{\theta_B}}{\cos{(\theta_B-\psi)}}\right) \, \left(\frac{\Lambda}{2\pi L}\right).
\end{equation}
For  the case, $\theta_B=2 \psi =\pi/6$, the selectivity is given by, 
\begin{equation}
\label{dtb}
\Delta \theta_{Bragg} \approx 0.7 \frac{\Lambda}{L}.
\end{equation}

Substituting Eq.~\ref{dtb} and Eq~\ref{lmt} into the inequality, Eq.~\ref{condition},  yields the minimum discernible number of waves of tilt,
\begin{equation}
b \ge 0.7 \frac{\Lambda}{\lambda}.
\end{equation}
Similarly, for an optical vortex the minimum discernible OAM over the aperture is
\begin{equation}
|l| \ge 4.5 \frac{\Lambda}{\lambda}.
\end{equation}
Finally, in order to differentiate an LG OAM state with a thick phase hologram, given with our estimate of $\Delta \theta_{Bragg}$ with  $\theta_B=2 \psi =\pi/6$, then we must ensure that,
\begin{equation}
|l| \ge 40.2 \left(\frac{\Lambda}{\lambda}\right)^2\, \left( \frac{w(z)}{L}\right)^2.
\end{equation}
If the grating spacing is on the order of a few wavelengths, then this requires that the emulsion thickness be substantially larger than  he waist of the beam. This constraint is within the parameters afforded by PTR glass for both LM and OAM photons. PTR glass has Bragg selectivities approaching a few waves of tilt for LM photons. The relaxation of this constraint by using larger quantum numbers will result in the requirement for larger apertures yielding an optimization problem.  A more detailed analysis is required.

\section{From CM theory to practical sorter}
\noindent
While the analysis presented here suggests that a high efficiency
single optical element sorter is feasible with commercially available
materials and holographic recording techniques, further work is
needed.  First, CM theory yields a solution with perfect efficiency and we know 
from other applications of PTR glass that it can yield high 
efficiencies for a few multiplexed gratings \cite{Ciapurin:06}. However, 
we need to examine the decrease in efficiency as a function of the 
number of multiplexed recordings for a fixed thickness of PTR glass.  
Second, the PTR glass is recorded in the UV spectrum and we will 
ordinarily use the hologram at another frequency.  For the thick
Bragg gratings discussed in Sec.~\ref{sec:CM} the wavelength scaling is trivial; however, the 
wavelength scaling for OAM states need to be addressed.  While this is not a problem for the thick Brag grating recordings discussed here; it nevertheless may be a serious issue for the extension of this work to qu$d$it photon states with both spatially varying amplitude and
phase functions. Finally the CM solution assumed perfect Bragg matching.  We need to further examine 
its sensitivity to alignment (linear and rotational) especially for  OAM photons as discussed in the last section. 

In this paper we examined the use of LM photons to expand the dimensionality of state space for QKD. A critical element is the design of a highly efficient sorter for LM MUB states. We believe we have made such a case and would like to draw attention for the need for  its experimental realization.  The beams emerging from a laser are ordinarily Gaussian beams,  however we considered here planar wavefronts with uniform intensity. It is customary for  experimentation to spatially filter and expand the waist of a Gaussian beam and to overfill an aperture $D$. The beam therefore closely approximates a uniform intensity wavefront and diffractive edge effects are minimal since $D \gg \lambda$.  The beam then is further attenuated to achieve the desired photon count. Notwithstanding this, substantial work as been done since the early 70's to analyze holograms produced with plane wavefronts with variable intensity profiles, e.g. Gaussian beams \cite{SJ:77, MGM:80, Kenan:78, Sidorovich:75}.  As long as the the grating strengths are large enough then high efficiencies can be obtained \cite{MGM:80}.  Since the grating strength is proportional to the waste of the beams our planar wavefronts here are consistent with the use of these more realistic beams \cite{SJ:77, MGM:80}.    It is clear that more work is needed for the non-planar helical wavefronts of photons with OAM.  

There continues to be active research in the efficient sorting of photons with OAM as evidenced by two papers appearing in the literature since this manuscript was posted \cite{Berkhout:10, Li:10}.  However, one of these approaches \cite{Berkhout:10} uses SLM's to transform the azimuthal position of the input into a transverse position at the output.  This device has (1) low efficiency, (2) requires multiple optical elements, and (3)  is not capable of sorting an arbitrary superposition state  of OAM at the single photon level.  While it can measure the relative fraction of a specific OAM superposition state for a beam of photons, it will not sort at the single photon level.   It is not a sorter appropriate for QKD with photons with OAM.  Similarly,  the mode analyzer (MODAN) described in \cite{Li:10}  requires a cascade of Mach-Zehnder interferometers to perform the fractional Fourier transformations.  It  is similar to the device reported by Leach et al. \cite{Leach:04} and is prohibitively difficult to stabilize.  The device reported here for LM states can achieve relatively high efficiencies in sorting single photons in a high-dimensional state space for applications to QKD and requires only a single optical element per $MUB$ basis.  While we are currently working on the design of an OAM volume-holographic grating, there appears to be no practical reason not to utilize the LM states.  In particular, one of the major obstacles in LM or OAM QKD as opposed to Polarization or Energy-Time QKD is due to the relative fragility of the states in propagation \cite{Paterson:05, Tyler:09}. However, the simplest form of adaptive optics is tip-tilt correction of the wavefront distortions as it propagates in free space--- a correction fundamental to LM QKD.  Perhaps a hybrid system based on the volume holographic system proposed here and the Berkhout et al. system \cite{Berkhout:10} would be an interesting area to explore. 

Experimentally, we are currently  producing and testing a
3-state k-QKD $MUB_1$ sorter. Our goal is to test its performance
using states generated by a single phase-modulated SLM \cite{Gruneisen:08}.  If
the MUB-state sorters described here can be produced, they should have
far more utility in quantum information processing than just QKD, e.g.
as an essential element in linear quantum computing \cite{Knill:01}.  

%\section*{Acknowledgements}
\begin{acknowledgements}
We wish to acknowledge important discussions with Leonid Glebov, Glen Tyler, Raymond Dymale and Mark Gruneisen.  We also acknowledge Robert Boyd for keeping us abreast of activities in this field by colleagues in  Scotland, Leiden and Kirtland AFB. We are also grateful for the advice provided by Grigoriy Kreymerman, William Rhodes, Angela Guzman and Chris Beetle of the FAU Quantum Optics group. We thank Anna Miller for reviewing this manuscript. This work was supported, in part, by a grant from the Quantum Information Group at the Air Force Research Laboratory (AFRL/RITC). 
\end{acknowledgements}

%\section*{References}

\end{document}